\documentclass{elsart}

\usepackage{epsfig,subfigure,latexsym,amsmath}
\usepackage{amssymb}
\usepackage{epsfig}
\usepackage{rotating}

\begin{document}

\input epsf

\begin{frontmatter}

\title{Adjustment studies in self-consistent relativistic mean-field models}
\author[lanl]{T. J. B{\"u}rvenich}
\author[lanl]{D. G. Madland}
\author[erl]{P.--G. Reinhard}
\address[lanl]{Theoretical Division, Los Alamos National Laboratory, Los Alamos, New Mexico 87545, USA}
\address[erl]{Institut f\"ur Theoretische Physik II, Universit\"at Erlangen-N\"urnberg,
Staudtstrasse 7, D-91058 Erlangen, Germany}

\begin{abstract}
We investigate the influence of the adjustment procedure and the
set of measured observables on the properties and predictive power of relativistic self-consistent
mean-field models for the nuclear ground state. These studies are performed with the point-coupling variant of the relativistic
mean-field model. We recommend optimal adjustment algorithms for the general two-part problem and we identify
various trends and dependencies as well as deficiencies of current models.
Consequences for model improvements are presented.
\end{abstract}

\end{frontmatter}

\section{Introduction}
There are many experimental data on the nuclear ground state. They include the nuclear mass, form factor, binding energy,
single-particle states, half lives,
etc. Moreover, in the future we can expect a dramatic increase in 
nuclear structure data due to new facilities that are ramping up or are
envisioned as, for example, the future GSI facility and RIA. Our ability to
understand the structure of the nuclear ground state is strongly connected to
formulating nuclear models that deliver simultaneously a quantitative description of all
measured ground-state observables.  Then, present (and future) experimental
data on nuclear ground states can be given a more realistic and complete physical
interpretation. Observables related to the mass, densities and geometries of
the nucleus as well as specific nuclear level data can be calculated with
microscopic mean-field models.

Modern self-consistent mean-field models like the Skyrme-Hartree-Fock (SHF) approach (for reviews see \cite{skyrme,review}) or the relativistic mean-field (RMF)
model (with finite-range mesons, RMF-FR, or point-couplings, RMF-PC), see Refs. \cite{review,Rei89,SW86,Ring,Nik92,HMMMNS94,Buer02,rusnak} have reached a mature status in nuclear structure physics.
They are phenomenology based and are designed to allow a description of nuclear ground states
throughout the nuclear chart, with very light systems ($A < 16$) being the exception. 
With few adjustable coupling constants (6-12) they deliver a plethora of microscopic information for a given nucleus. 
Their modern reinterpretation as Kohn-Sham schemes in the spirit of density functional theory \cite{Drei90}
explains their success. They can be viewed as effective field theories for
nucleonic degrees of freedom below an energy scale of $\Lambda \approx 1$~GeV. 
Short distance physics is absorbed into the various (contact-) terms and coupling constants.

Independent of the interpretation of these models, each term in the Lagrangian or energy functional
introduces a parameter or coupling constant that (yet) cannot be determined from more fundamental theories. Thus, the models
require parameter adjustment. A good functional form with a bad fit has small predictive power, thus the adjustment procedure is critical. 
For the description of exotic nuclei and reliable extrapolations to the drip-lines, a well adjusted mean-field is as important
as a proper treatment of the pairing correlations and/or a possible beyond-mean-field treatment of ground-state correlations and
excited states. 

Even though the models have reached high accuracy for various observables (we will discuss this in more detail later), they still
exhibit significant short-comings and puzzles. Some of these are: a) the asymmetry energy of $a_4 \approx 38$~MeV in RMF models appears
to be too large compared to other accepted empirical values of $a_4 \approx 32~$MeV, b) compared to SHF, the saturation density in RMF is smaller, the energy at
saturation larger, c) there are erroneous mass and isovector trends in binding energies along isotopic and isotonic chains, d) surface thicknesses are too
small (up to 5 \%), e) the compressibility $K$ seems to be determined in the RMF-PC variant ($K \approx 260$~MeV), while RMF-FR forces with values
between $172-280$~MeV exhibit roughly comparable predictive power for ground-state properties, f) axial fission barriers for actinides and superheavy nuclei tend to be
systematically lower for RMF compared to SHF, g) within RMF the neutron matter equation of state is too stiff for high densities, h) pseudo-spin splittings in 
RMF are too large \cite{pseudosplitting}, and i) the available types of ground-state observables are (yet) not sufficient to uniquely determine
all of the coupling constants.
Note that the asymmetry energy of RMF models with density-dependent coupling constants is about $\approx 33$ MeV \cite{Vret}.
For a more detailed description of these short-comings, see Refs. \cite{review,Rei89,Buer02,fissbar}.

In this paper we study the influence of certain methods of parameter adjustment on the resulting model, and its sensitivity to
selection of observables and weights. The consequences and issues related to the adjustment of phenomenological mean-field models
are valid for a variety of other models and applications.  

We first describe the model with which we perform our studies. We then discuss different adjustment protocols.
This is followed by specification of the possible sets of measured observables and their weights.
The influence of these two topics on the resulting effective forces is then presented and discussed.

\section{Description of the Model}
For our studies we employ the point-coupling variant of the
RMF model. Previous studies \cite{Nik92,HMMMNS94,Buer02} have shown that - with differences in detail - this
model performs very similar to the conventional RMF-FR approach with
finite-range fields. Yet, except for density profiles, which in the RMF-PC are
smoother and resemble the ones obtained in SHF calculations, little
sensitivity of observables with respect to mesons versus point couplings and
derivatives has been found. In Ref. \cite{review}, the performance of various self-consistent models,
including the point-coupling approach, are discussed. Our current point-coupling Lagrangian is given by 
\begin{equation}
\begin{array}{lcl}
  {\mathcal L} 
  & = & 
  {\mathcal L}^{\rm free} + {\mathcal L}^{\rm 4f} + {\mathcal L}^{\rm hot}
  + {\mathcal L}^{\rm der} + {\mathcal L}^{\rm em},~~~{\rm where}
\\[12pt]
  {\mathcal L}^{\rm free} \hfill
  & = & 
  \bar\psi ({\rm i}\gamma_\mu\partial^\mu -m)\psi,
\\[6pt]
  {\mathcal L}^{\rm 4f} \hfill
  & = & 
  - \tfrac{1}{2}\, \alpha_{\rm S} (\bar\psi\psi)(\bar\psi\psi)
  - \tfrac{1}{2}\, 
    \alpha_{\rm V}(\bar\psi\gamma_\mu\psi)(\bar\psi\gamma^\mu\psi)
\\[6pt]
  & &
   - \tfrac{1}{2}\, \alpha_{\rm TS} (\bar\psi\vec\tau\psi) \cdot
   (\bar\psi\vec\tau\psi)
  - \tfrac{1}{2}\,  \alpha_{\rm TV} (\bar\psi\vec\tau\gamma_\mu\psi)
    \cdot (\bar\psi\vec\tau\gamma^\mu\psi),
\\[6pt]
  {\mathcal L}^{\rm hot} 
  & = &  
  - \tfrac{1}{3}\, \beta_{\rm S} (\bar\psi\psi)^3 - \tfrac{1}{4}\, 
    \gamma_{\rm S} (\bar\psi\psi)^4 - \tfrac{1}{4}\, \gamma_{\rm V} 
    [(\bar\psi\gamma_\mu\psi)(\bar\psi\gamma^\mu\psi)]^2,
\\[6pt]
  {\mathcal L}^{\rm der} 
  & = & 
  - \tfrac{1}{2}\,\delta_{\rm S}(\partial_\nu\bar\psi\psi)
    (\partial^\nu\bar\psi\psi)  
  - \tfrac{1}{2}\,  \delta_{\rm V} (\partial_\nu\bar\psi\gamma_\mu\psi)
    (\partial^\nu\bar\psi\gamma^\mu\psi)
\\[6pt]
  & &
   - \tfrac{1}{2}\, \delta_{\rm TS} (\partial_\nu\bar\psi\vec\tau\psi) \cdot
   (\partial^\nu\bar\psi\vec\tau\psi)
  - \tfrac{1}{2}\, \delta_{\rm TV} (\partial_\nu\bar\psi\vec\tau\gamma_\mu\psi)
    \cdot (\partial^\nu\bar\psi\vec\tau\gamma^\mu\psi),
\\[6pt]
  {\mathcal L}^{\rm em} 
  & = & 
  -  e A_\mu\bar\psi[(1-\tau_3)/2]\gamma^\mu\psi -  
    \tfrac{1}{4}\, F_{\mu\nu} F^{\mu\nu}.
\end{array}
\label{eq:lagrang}
\end{equation}
We employ BCS+$\delta$-force pairing with a smooth cutoff given by a Fermi function in the single-particle
energies \cite{pair}. In each adjustment, the pairing strengths are adjusted simultaneously with the
mean-field parameters. There are 11 coupling constants in Eq. (\ref{eq:lagrang}) and a total of 13 coupling constants -- including the pairing strengths for protons and neutrons -- to be determined. This model has been employed to adjust the RMF-PC force PC-F1 \cite{Buer02}. Based on experience, it contains a minimal
number of terms necessary for a description of nuclear ground states comparable to established RMF-FR models.

We note that only the direct Coulomb term is employed. We have studied the
Coulomb exchange in Slater approximation (as it is used in SHF calculations) and have learned that its effect constitutes
a rather small correction for observables such as binding energies and
radii. Its effect can be balanced by a rearrangement (readjustment) of the mean-field coupling constants. 
This does not, however, exclude the necessity for more general Coulomb exchange-correlation development in future 
models. 
\section{Adjustment Protocols and Strategies}
The adjustment of mean-field models is a non-straightforward and biased endeavor. First of all, there is no {\em canonical} way to do it.
One has the freedom of choosing various kinds of observables and nuclei and of choosing the various uncertainties that enter the
$\chi^2$ for the least-squares adjustment. Whereas some choices are obvious as, for example, binding energy as the most accurately measured
ground-state observable, additional observables are not as obvious.
Possible choices include rms radii, diffraction radii, surface thicknesses, neutron radii, spin-orbit splittings, 'pseudo observables'
like nuclear matter properties, and more.
Each parameter adjustment is a balancing act between different types of observables, where the balance is influenced by the properties
of the model and the weights for the various observable types. As we will see, current mean-field models exhibit a competition between
the binding energy and the form factor-related observables in the fit. The nuclei chosen are usually spherical nuclei. Adjustments for
deformed nuclei are more difficult.

The principle problem in the adjustment of the model parameters arises from the complicated structure of the $\chi^2$ hyper-surface. It usually exhibits several separated local minima and
its complicated structure results from its dimension of $D \approx 6-12$ and the nonlinear dependence of the observables on the coupling constants.
Often, different minima differ by the ability of the model to describe certain kinds of observables (energy, radii, ...) better than others. Also, we have sometimes found
minima with quite low $\chi^2$ values and a good description of the
fit data, but with less predictive power for the same observables in other nuclei. We interpret these cases as an indication
that our interaction Lagrangian ${\mathcal{L}}$ is not (yet) complete.

Changing the adjustment protocol, i.e., the choice of observables (see next section for a detailed description) changes the
parameters, but this solution will still remain close to the point where it started, or, put differently, the relative minimum
will still be the same when one uses down-hill methods which - in each step - try to lower $\chi^2$. On the other hand,  Monte-Carlo methods like Simulated
Annealing (SA) or Genetic Algorithms (GA) allow one to leave a certain region in parameter space and explore the surrounding region. Simulated Annealing
explores the $\chi^2$ landscape by accepting -- with a certain probability that depends on an external temperature -- parameter configurations
that increase $\chi^2$. Thus, local minima can be left and ridges can be overcome. With decreasing temperature, the probability for
accepting higher $\chi^2$ configurations decreases, so that the algorithm slowly settles down in a minimum. Genetic algorithms obtain their
ability to roam the parameter space by recombining different parameter vectors, their mutation and then selection of the best solutions
in terms of {\em fitness}. GA can also be combined with {\em local optimization} with down-hill algorithms such as the ones described below. Our experience 
to date is that for new forces, a Monte-Carlo type search is mandatory for a first exploration of parameter space. Once several regions around minima
have been identified, down-hill algorithms can yield the minimal $\chi^2$ values. 
The general optimization is thus a two-part problem.
Since in the present study we are concerned with variations of adjustments starting
in the same local minimum, we have employed only down-hill methods here.

A variety of commercial and non-commercial optimization routines is available on the market. Examples for downhill algorithms are simple grid searches, Powell's
conjugate-directions method, and Leven\-berg - Marquardt \cite{NumRec,Bevington}. 
The grid-search method, which in every iteration changes each parameter separately before comparing the new $\chi^2$ value with the 
previous one is quite inefficient due to large parameter correlations in these models (for discussion of these correlations see Subsection \ref{parameters}). 
A better approach is Powell's method, which starts out like a grid search but adjusts the directions in parameter space
to optimize the down-hill motion. Our experience is, however, that in  the hilly $\chi^2$ hyper-surfaces this method often gets
stuck and progresses very slowly. 
Levenberg-Marquardt combines the simple gradient descent with a parabolic expansion of the $\chi^2$ hyper-surface around the 
minimum. It dynamically switches between these two scenarios: far from the minimum, the gradient descent is preferable. Closer to the minimum, where
the gradient descent would only asymptotically reach the bottom, the parabolic expansion is used iteratively. An additional 
method, denoted as {\em trial step},  uses the information of already obtained $\chi^2$ values for a parabolic interpolation between them. The minimum of the
parabola is the new guess for $\chi^2$.

Based on previous \cite{Buer02} and present experience \cite{fits}, Levenberg-Marquardt plus the trial step in parameter space
\cite{Bevington}
(henceforth denoted as {\em Bevington curved step}) has been chosen here. But this algorithm 
needs to be restarted  several times to reach the minimum in the $\chi^2$ hyper-surface.

Since the $\chi^2$ landscape of these models appears to resemble a hilly landscape, a combination of simulated annealing and diffusion Monte-Carlo \cite{CompPhys} has been
tested in \cite{Buer02}. This method found the region containing the parameter set PC-F1, and Bevington curved step led
to the actual minimum containing PC-F1.
Algorithms and techniques such as SA, GA, Evolution Strategies, and Neural Networks
for the optimization of $\chi^2$ might become more important with the increasing complexity of nuclear density functionals.  They could be used
for the optimization of the coupling constants and/or the optimization of basis terms in the expansion of the energy functional.

The task of parameter adjustment is simplified if the adjustment is initialized with realistic parameter values,
i.e., good first guesses. These initial values could be obtained from or related to more fundamental theories like QCD.
In the future, as the number of terms and observables increase, such information will become
increasingly important. One guidance has emerged already: naive dimensional analysis (NDA) or QCD scaling,
which examines scaled coupling constants obtained from the original coupling constants by scaling them with the pion decay constant, $f_\pi = 93$~MeV, and
the QCD mass scale, $\Lambda \approx 600-1000$~MeV \cite{We90,MG84,We79,Fri96}. These scaled and dimensionless coupling constants are called {\em natural} if they
are of order 1. So far, QCD scaling has worked well for recent sets of coupling constants with excellent predictive power \cite{Buer02,rusnak,Fri96,Serot1}.
We expect that QCD scaling will provide further useful guidance for upcoming more detailed model development.

A further connection between relativistic point-coupling models and QCD sum rules, the chiral condensate, and pion dynamics was drawn in Refs. \cite{finelli1,finelli2,vretenar1} where the
approximate values of the coupling constants are related to the physical properties of low-momentum QCD.
Yet, due to truncated model spaces and other approximations, fine tuning will probably always be necessary to obtain optimal predictive power.

\section{Sets of Measured Observables and Weights}
The sets of observables and weights used for the adjustment of the RMF-FR force NL-Z2 \cite{NLZ2} and the point-coupling force
PC-F1 are shown in Table \ref{nuc}. These weights are based upon expectation and experience and
constitute an average compromise between the actual measurement uncertainties and
the requirement that the type of observable have a non-negligible influence in the $\chi^2$ minimization
or adjustment procedure. The 0.2\% weight on the experimental masses is greater than the corresponding experimental
uncertainties of 1.2 $\times$ 10$^{-3}$~\% (average) and the 0.5 \% weight on the experimental mean square charge radii is greater than the
corresponding experimental uncertainties of 0.17 \% (average).

Due to this choice of weights, the absolute value of $\chi^2$ or $\chi^2/{\rm dof}$ has only a relative meaning.
However, the comparison of different
$\chi^2$ values from various fits still indicates their relative goodness with respect to each other.  
There are further indications for the goodness of the fit, though.
The $\chi^2/{\rm dof}$ of PC-F1, for example, has a value of 2.75, indicating that the model does not fulfill our
expectations of performance reflected in our choice of weights. Furthermore, the correlated errors of the coupling constants,
which also depend on our choice of weights in the total $\chi^2$, still deliver relative information on how well the various
parameters have been determined. Due to strong parameter correlations (see Subsection \ref{parameters}) these correlated errors
range from 5-70 \% and, in the worst cases, are larger than the value of the coupling constant, indicating that a readjustment without that parameter
could, in principle, reach the same accuracy (see Ref. \cite{Buer02} for more details). 

We have separately reduced the weights of the observables entering our total $\chi^2$, keeping the remaining weights fixed, to
test the sensitivity of $\chi^2$ with respect to their choices. A 50 \% reduction of the weight for binding energy (0.2 \% $\rightarrow$  0.1 \%), diffraction radius 
(0.5 \% $\rightarrow$  0.25 \%), surface thickness (1.5 \% $\rightarrow$  0.75 \%), and rms radius (0.5 \% $\rightarrow$  0.25 \%)
leads to an increase of the total $\chi^2$ by factors of 2.9, 1.4, 1.2, and  1.5, respectively. The original total $\chi^2$ 
of PC-F1 is 99. As expected, these factors are all greater than unity. The fact that they are (approximately) uniformly
greater than unity indicates that (approximately) equal influence exists on the (final) interaction Lagrangian for
each of the four observable types. This was the goal in the choice of weights, but is by no means unique and, above all,
the choice of weights is primarily subject to the constraints of the actual (average) experimental errors in the
four observable types. Repeating this exercise, but with a full parameter search for each of the four cases, should
lead to to the same conclusion as long as one remains in the same minimum on the $\chi^2$ hyper-surface.

The set PC-F1 has been used as a basis for the majority of investigations in this study. It has been modified for
several runs to test the sensitivity of the resultant force with respect to various choices of observables as well as weights.
The nuclei chosen are either spherically magic or doubly-magic systems. 
These systems are reasonably good mean-field nuclei, where ground-state correlations play a minor role and they require minimal computational time. Deformed nuclei increase the computational effort and require considerations of vibrational-rotational correlations, which
are sometimes parametrized by phenomenological terms. However, these systems deliver additional information and hence will be considered
in future relativistic mean-field model refinements, as has already been seen in non-relativistic Skyrme-Hartree-Fock mass formulas \cite{HFMass3,HFMass4,HFMass5}.

\begin{table}[htb]
\begin{center}
\begin{tabular}{@{\hspace{0.1cm}}l@{\hspace{0.2cm}}c@{\hspace{3.0mm}}c@
{\hspace{3.0mm}}c@{\hspace{3.0mm}}c@{\hspace{3.0mm}}c@{\hspace{3.0mm}}c@
{\hspace{3.0mm}}c@{\hspace{3.0mm}}c@{\hspace{3.0mm}}c@{\hspace{3.0mm}}c@
{\hspace{3.0mm}}c@{\hspace{3.0mm}}c@{\hspace{3.0mm}}c@{\hspace{3.0mm}}c@
{\hspace{3.0mm}}c@{\hspace{3.0mm}}c@{\hspace{3.0mm}}c@{\hspace{3.0mm}}c@
{\hspace{0.1cm}}}
observable & weight &  \begin{sideways}$^{16}$O\end{sideways}  &  
\begin{sideways}$^{40}$Ca\end{sideways}  &  \begin{sideways}$^{48}$Ca
\end{sideways}  &  \begin{sideways}$^{56}$Ni\end{sideways}  & \begin{sideways} 
$^{58}$Ni\end{sideways}  & \begin{sideways} $^{88}$Sr\end{sideways}  &  
\begin{sideways}$^{90}$Zr\end{sideways}  &  \begin{sideways}$^{100}$Sn
\end{sideways}  &  \begin{sideways}$^{112}$Sn\end{sideways}  & 
\begin{sideways} $^{120}$Sn\end{sideways}  & \begin{sideways} $^{124}$Sn
\end{sideways}  &  \begin{sideways}$^{132}$Sn\end{sideways}  &  
\begin{sideways}$^{136}$Xe\end{sideways}  & \begin{sideways} $^{144}$Sm
\end{sideways}  & \begin{sideways} $^{202}$Pb \end{sideways}  &  
\begin{sideways}$^{208}$Pb\end{sideways}  & \begin{sideways} $^{214}$Pb
\end{sideways}  \\ \hline
$E_{\rm B}$ & 0.2~\% &+&+&+&+&+&+&+&+&+&+&+&+&+&+&--&+&+\\
$R_{\mbox{\scriptsize diff}}$ & 0.5~\% &+&+&+&--&+&+&+&--&+&+&+&--&--&--&--&+&--\\
$\sigma$ & 1.5~\% &+&+&+&--&--&--&+&--&--&--&--&--&--&--&--&+&--\\
$r_{\mbox{\scriptsize rms}}^{\mbox{\scriptsize ch}}$ & 0.5~\% &--&+&+&+&+&+&+&
--&+&--&+&--&--&--&+&+&+\\ \hline
$\Delta_{\mbox{ \scriptsize p}}$& 0.05~MeV & - & - & - & - & - & - & - & - & - 
& - & - & - &
+ & + & - & - & -\\
$\Delta_{\mbox{ \scriptsize n}}$& 0.05~MeV & - & - & - & - & - & - & - & - 
& + & + & + & - &
- & - & - & - & -\\[20pt]
\end{tabular}
\begin{tabular}{lcl}
\hline
 PC-F1  &:&   identical to above set\\
 PC-masses &:&  binding energies only\\
 PC-asym &:& additional constraint on symmetry energy $a_{\rm sym}=34$ MeV\\
 PC-surf &:&  reduced weights on $\sigma$ = 0.025\%\\
\hline
\end{tabular}
\end{center}
\caption{
Upper part:
Observables and chosen weights of nuclei used for
the Bevington curved step fitting procedure.  $E_B$ denotes the binding energy, $r_{\mbox{\scriptsize rms}}^{\mbox{\scriptsize ch}}$ is the
(charge) rms radius, $R_{\rm diffr.}$ denotes the
diffraction radius, $\sigma$ the surface thickness, and $\Delta_{\rm
p}$ and $\Delta_{\rm n}$ are the proton and neutron pairing gaps.  A
$+$ indicates an observable contributing to the total $\chi^2$. For
the actual experimental values see Ref. \protect\cite{NLZ2}.
Lower part: Variation of the sets for the four cases 
discussed in the present work.
}\label{nuc}
\end{table}
\begin{figure*}[htb!]
\centerline{\epsfxsize=13cm \epsfbox{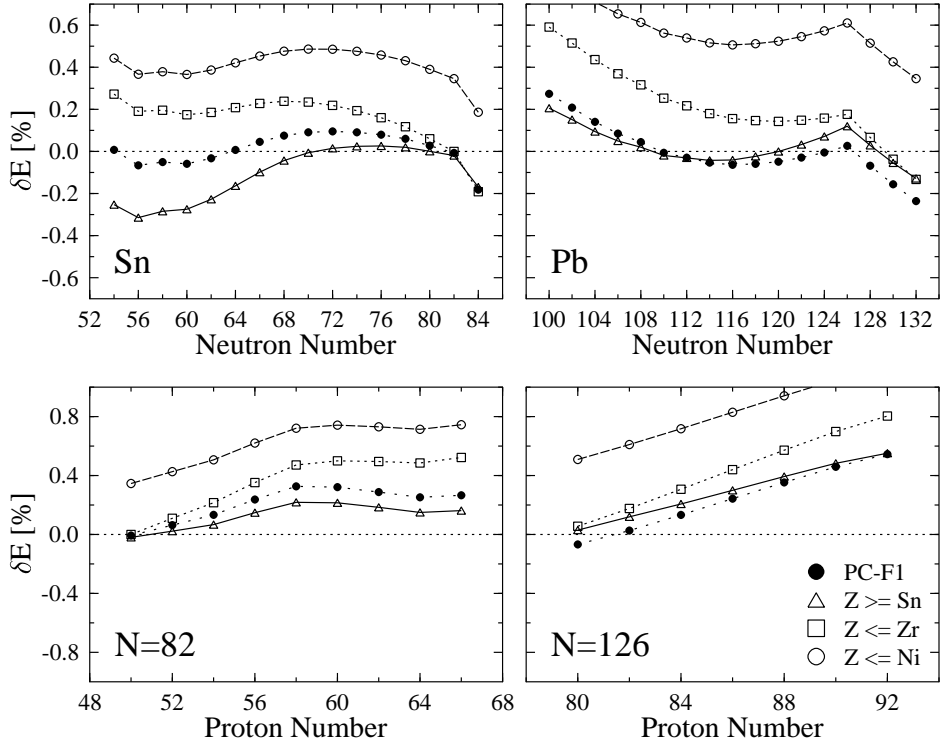}}
\caption{\label{fig:varnuclei}Relative errors of binding energies for selected isotope (top) and isotone (bottom) chains for different 
subsets of nuclei from Table \protect\ref{nuc}. 
}
\end{figure*}

We have tested the sensitivity of the RMF-PC model with respect to the mass range of nuclei included
in the set of observables shown in Table \ref{nuc}. 
The calculations displayed in Fig. \ref{fig:varnuclei} indicate a trade-off
between light and heavy systems in adjustment procedures
including nuclei ranging from oxygen to lead isotopes. The lighter the heaviest nuclei
in the adjustment procedure are, the stronger the over-binding of the
heavier species. A possible solution might be phenomenological terms
accounting for correlations and Wigner energy and/or density dependent or gradient terms.
Enhanced new-generation mean-field models that
constitute a better approximation to the exact functional would
presumably remedy this problem. A more
pragmatic approach is to adjust forces for light and heavy systems
separately, as done with TM2 (light systems) and TM1 (heavy
systems) \cite{Sug94}, but we do not pursue such a course here. Also, Hartree-Fock mass formulas focus only on nuclear masses.
The approach here applies to 'complete forces' designed to describe all types of
observables simultaneously. 
\section{Results}
Certain problems in the adjustment procedure presumably arise from missing ingredients in the models. It is not yet clear how well correlation
energies can be absorbed in the energy functional. Pairing correlations like $T=0$ pairing and pairing-mean-field couplings 
are usually missing. More generally, in both the isovector and isoscalar channels, terms might be missing that lead to deficiencies in, for example, the density dependence.
The missing ingredients are absorbed in a somewhat uncontrolled manner in the coupling constants during the adjustment
procedure. This, in turn, results in an entanglement of different parts of the effective interaction, which can then bias
its physical interpretation.
\subsection{Parameter Correlations}
\label{parameters}
\begin{figure*}[htb!]
\centerline{\epsfxsize=12cm \epsfbox{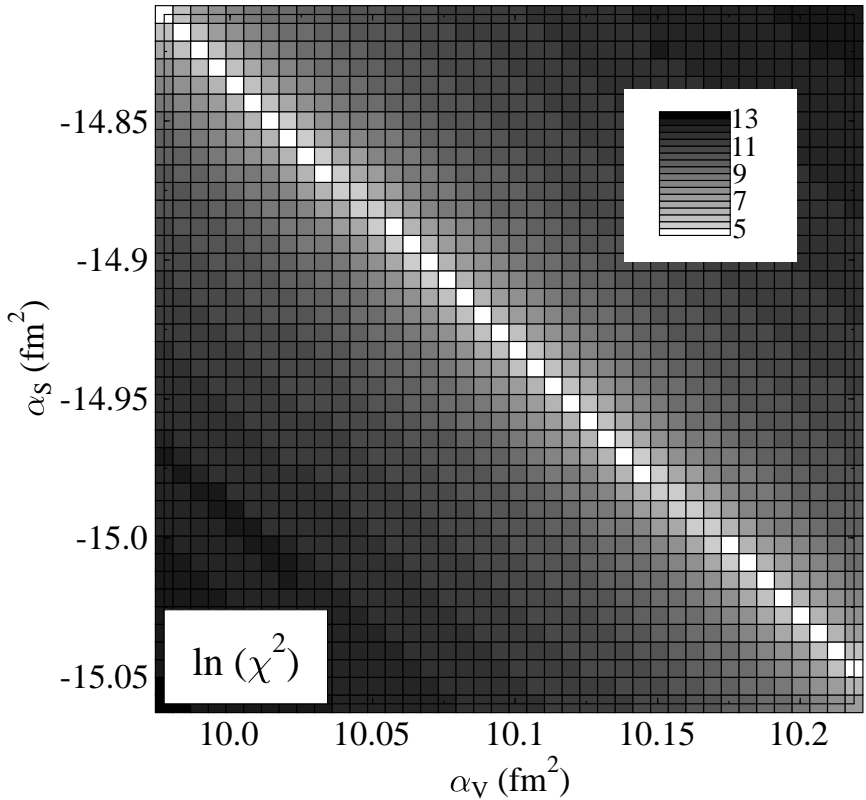}}
\caption{\label{fig:corr1}Natural logarithm of the total $\chi^2$  for different combinations of the isoscalar coupling constants
$\alpha_{S}$ and $\alpha_{V}$ (here we use units of fm instead of MeV). 
}
\end{figure*}

A well-known problem that occurs when adjusting mean-field models is the presence of strong
parameter-correlations.  Figure \ref{fig:corr1} shows as an example the total
$\chi^2$ as a function of the RMF-PC coupling constants $\alpha_S$ and
$\alpha_V$, all other parameters having been fixed. These two coupling constants
are responsible for the bulk part of the nuclear potential, and there exist
correlations for the equivalent coupling constants in the RMF-FR and the SHF
approach. The reason for these correlations is that not their individual values, but rather
combinations of them have physical significance, namely, the sum $\alpha_S +
\alpha_V$ determines the isoscalar part of the nuclear potential, while their
difference $\alpha_S - \alpha_V$ determines the size of the spin-orbit
potential. Thus the same potential depth can be achieved by various values of
$\alpha_S$ and $\alpha_V$ as long as their sum amounts to $V\approx 60$~MeV.
This explains the strong linear correlation. The limitations to this freedom
arise from the density dependence of the isoscalar fields and the size of the spin-orbit
potential which appear for larger deviations from the values in
PC-F1. It is curious, though, that the linear correlations are not
constant along that line in $\alpha_S-\alpha_V$ space, but that there is
rather a ridge which rises and falls (at least) twice. This hints at further nonlinear effects
associated with certain parameter combinations.
These correlations have led Rusnak and Furnstahl \cite{rusnak} to replace the original coupling constants by 
certain linear combinations of them.
In the present study, however, we use the original parameters in order to compare to our previous adjustments. 

\subsection{Weights and asymmetry energy}
\begin{figure*}[htb!]
\centerline{\epsfxsize=15cm \epsfbox{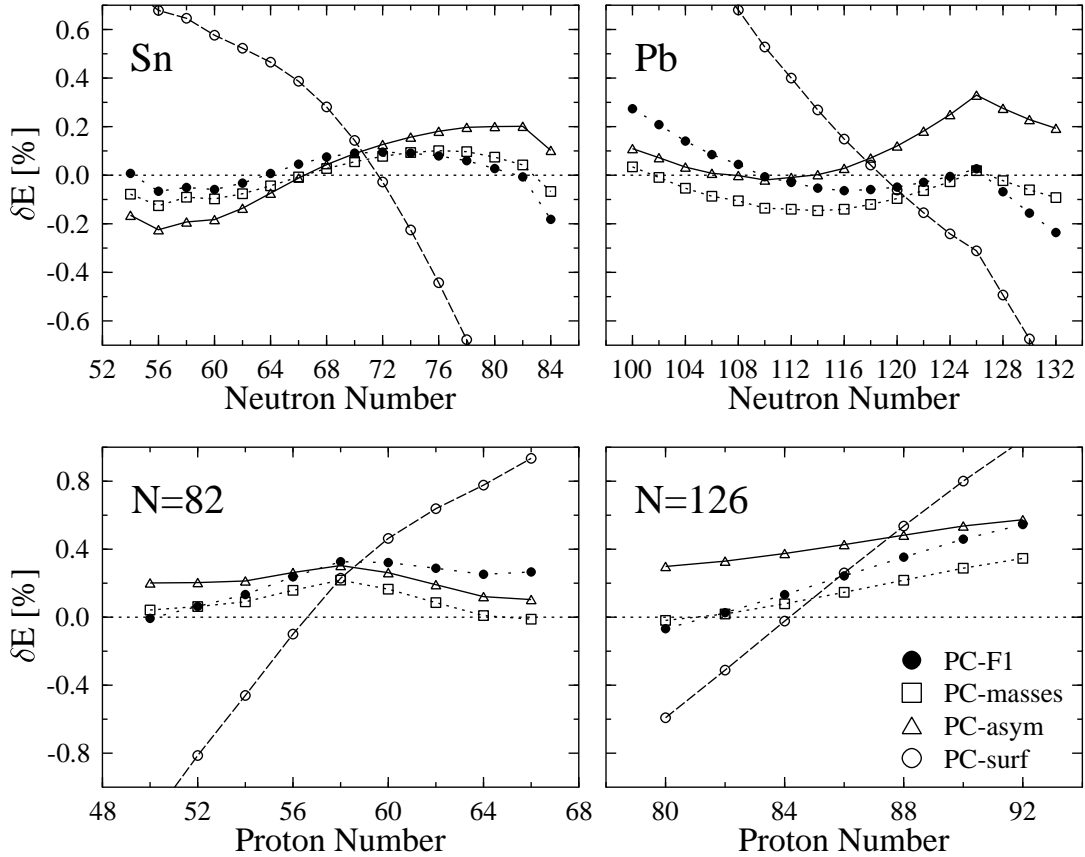}}
\caption{\label{fig:weights}Relative errors of binding energies for selected isotope (top) and 
isotone (bottom) chains for different choices of observables and weights, and a constraint on $a_4$. 
}
\end{figure*}

Here, we study isotope and isotone chains and the impact of a variation in weights 
as well as a constraint on the value of $a_4$ (see Fig. \ref{fig:weights}).
In these isotope and isotone
chains, wrong trends generally indicate a problem of balance between
isoscalar and isovector fields (for isotones, also the
Coulomb field).
For example: a neutron added to a nucleus feels
isoscalar attraction but isovector repulsion; a proton added to the
nucleus feels isoscalar attraction, isovector attraction (since there
are more neutrons than protons except for $N=Z$ or proton-rich nuclei)
and repulsion due to the electromagnetic field.  If these different
fields are not balanced, an erroneous trend results. So, what we see is
the sum of wrong trends with mass number, isospin, and electric
charge. Furthermore, the
distribution of errors cannot be explained by missing ground-state
correlations. These would produce an arch between magic corner
points. Thus the trends in errors are genuine mismatches in the mean field. 

We can learn several things from these different adjustments. When adjusting only the
binding energies the trends in isotopes, but even more in isotones, are improved. Here
we have an indication that energy and form factor compete during the adjustment.
It is interesting that in the lead isotopes beyond $N=126$ the
trend seems to be reversed a bit and extrapolations might be more
reliable than with PC-F1. 

Including $a_4$ in the least-squares adjustment with a
fixed value of 34 MeV (the usual value in RMF is around 38 MeV) worsens
the isotope trends.  The isotone trends are improved but the offset
increases, and the nuclei become more over-bound. This may be attributed to
missing repulsion coming from a weaker isovector channel. 
Other adjustments with still smaller and probably more realistic values of $a_4$ (30-32 MeV) were performed, but
the resulting forces showed poor convergence for most non-magic nuclei.
 
Putting extreme emphasis on the
surface thickness $\sigma$ leads to very poor predictions for the
energies: for isotopes, a strong trend to under-binding results with
increasing neutron number; for isotones, a strong trend to over-binding
with increasing proton number results.

The isovector repulsion drives
the neutrons out. RMF models deliver a too small surface thickness and
a too large $a_4$. We can speculate that the too large $a_4$ value in
RMF models (both finite range and point-coupling variants) results
from some missing repulsion and/or wrong density dependence in the
isoscalar channel.  A smaller value of $a_4$ results in less repulsion
for neutrons, which reflects itself in the trend to over-binding in the
isotope chains. In this way we can also understand the case with strong
emphasis on $\sigma$: since the outer part of the surface thickness is
governed by the neutrons, enforcing large $\sigma$ leads to a very
large value of $a_4$ of $40.6$ MeV (see Table \ref{tab1}). Thus, for
masses, neutrons suffer too much repulsion (trend to under-binding in
isotopes), and protons suffer too much isovector attraction (see the
isotones). This finding supports the result for the constrained
$a_4$ fit.
\begin{figure*}[htb!]
\centerline{\epsfxsize=15cm \epsfbox{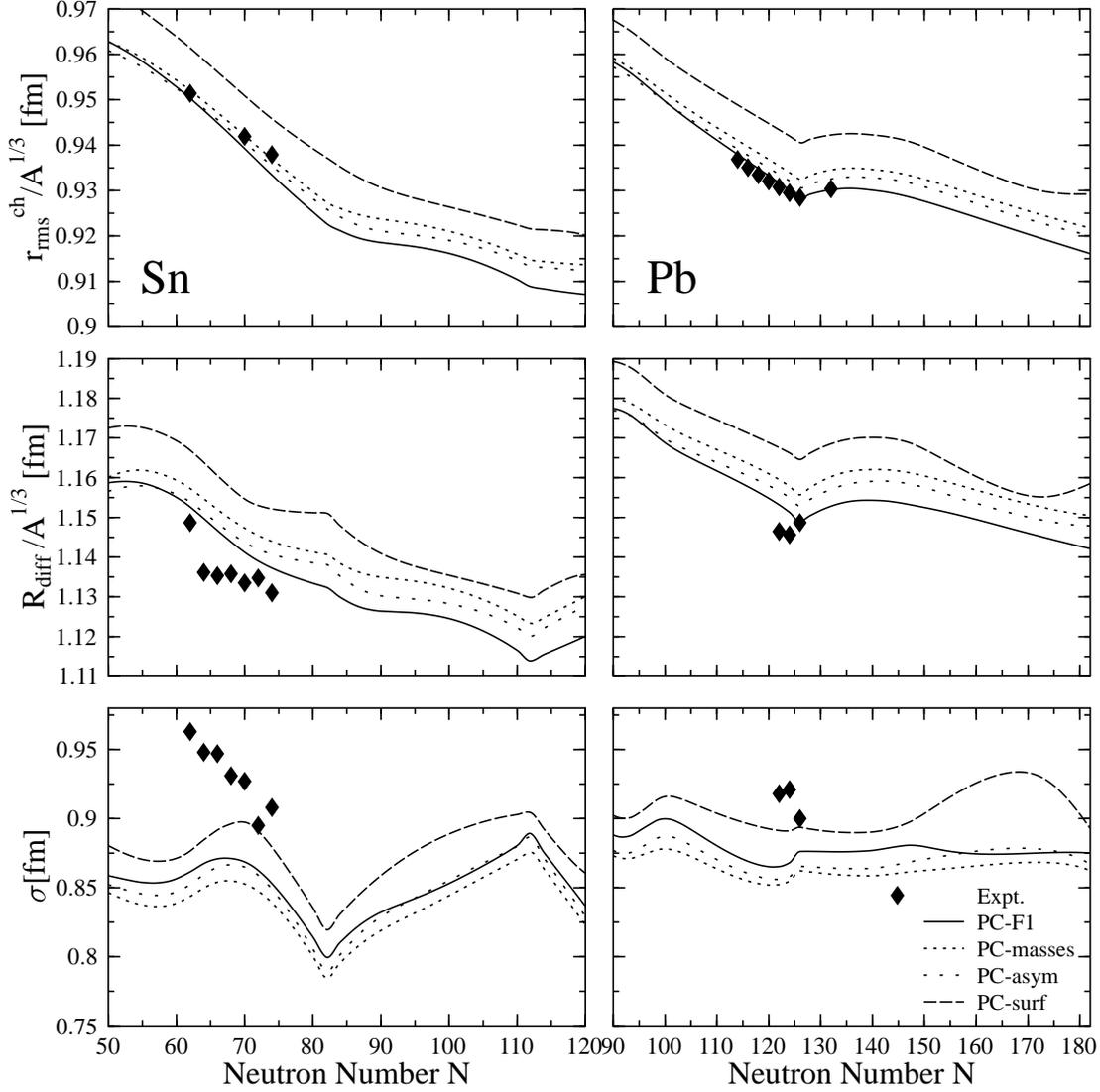}}
\caption{\label{fig:weightsform}Form factor observables for tin and lead isotopes for different choices of
observables and weights, and a constraint on $a_4$. 
}
\end{figure*}

\begin{figure*}[htb!]
\centerline{\epsfxsize=15cm \epsfbox{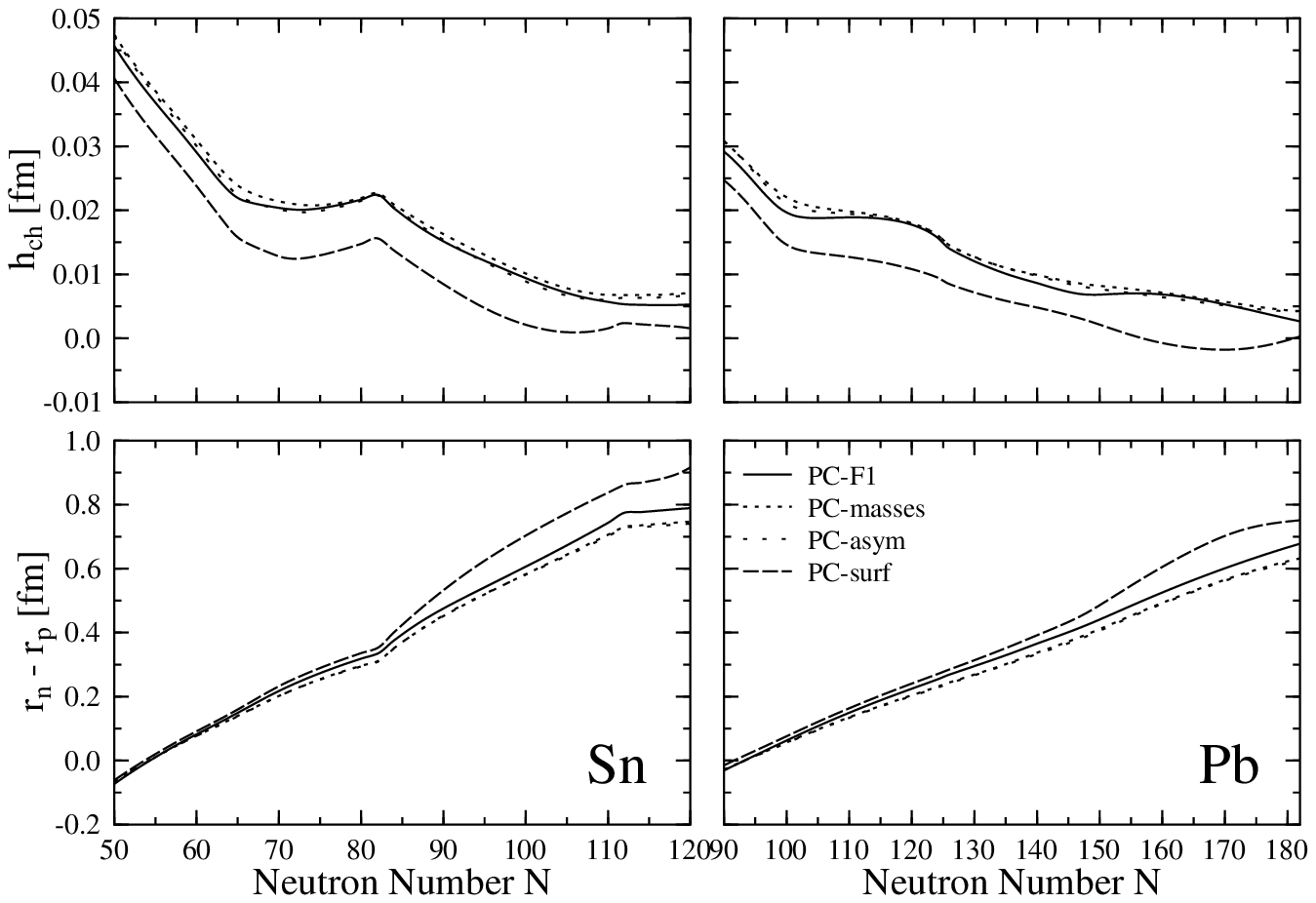}}
\caption{\label{skins}Halo parameters (top) and neutron skins (bottom) for tin and lead isotopes with the observables and
weights as indicated.
}
\end{figure*}

Figure \ref{fig:weightsform} illustrates the analysis of the form factors:
adjusting only to the binding energies (masses), the model tends to larger radii and
smaller surface thicknesses compared to PC-F1. A similar but not so strong
effect is seen in the fit enforcing $a_4 = 34$~MeV. Again, compared to PC-F1,
radii are larger and $\sigma$ is a bit smaller. Enforcing an extremely high
accuracy on $\sigma$ lifts it up to slightly more realistic values
(there is, however, still a trend to underestimation), but at the same time
leads to overestimation of the radii. This supports the hypothesis that the
isovector repulsion is replacing repulsion that should come from the
isoscalar channel. 

In Figure \ref{skins} we show two more quantities that
yield information on the surface region of the nucleus, namely the (neutron)
skin $r_{\rm n} - r_{\rm p}$ (difference of neutron and proton rms radii) and
the halo parameter defined as $h_{\rm ch}=r^{\rm ch}_{\rm rms}-\sqrt{0.6R_{\rm
diff}^2+3\sigma}$ \cite{Miz00a}. The most striking observation here is the
effect of the focusing on $\sigma$.  While
the neutron skin is slightly enhanced we see a strong decrease of the halo.  This means that while the neutrons as a whole move outward, the
very-low density region of the nucleus shrinks.  In the cases of adjustments
to masses only and the constrained $a_4$, the differences with PC-F1
are much smaller, the skins get slightly smaller while the halo increases by a
small amount. Again, we find an anti-correlation between halo and
skin.
\begin{figure*}[htb!]
\centerline{\epsfxsize=16cm \epsfbox{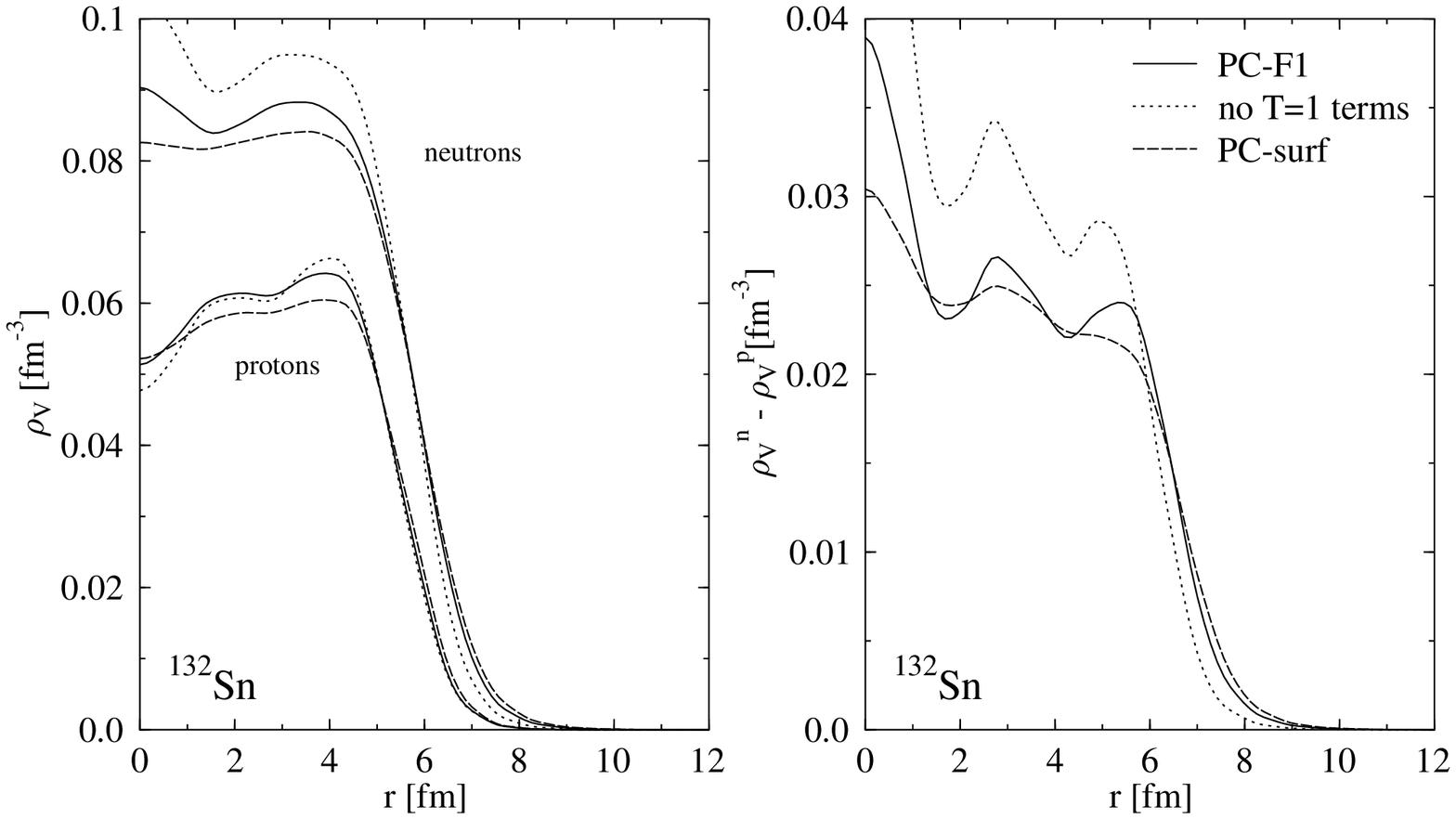}}
\caption{\label{fig:tin}Proton and neutron point densities (left) and their differences (corresponding to $\rho_{TV}$) for different choices 
of observables and weights with respect to the isovector channel and surface thickness $\sigma$. 
}
\end{figure*}

Figure \ref{fig:tin} shows the effects of the isovector terms
and the emphasis on $\sigma$ on the proton and neutron
densities. The main effects show up for the neutrons while the protons
remain much less affected. The neutrons react strongly to the absence of isovector terms, they are pulled back
into the interior of the nucleus and their contribution to the surface of the nucleus decreases. The emphasis
on $\sigma$ in the adjustment pushes them out.
Thus, any correlation between the neutron
radius and other observables is at the same time a correlation between
the surface thickness and these same observables.
\subsection{Nuclear matter}
The bulk properties of nuclear matter for the observables and weights discussed above are summarized in
Table \ref{tab1}.  
The effective mass ($m^*/m$) is insensitive to the four adjustments,
which means that the absolute magnitude of the scalar potential does not vary much.

The exclusion of geometric observables from the adjustment leads to
an $a_4$ value which is in better agreement with empirical data. This is
supported by the adjustment with strong emphasis on the the surface - $a_4$ is
driven to $40.6$ MeV. We also observed that selections of different nuclei do not influence $a_4$ 
drastically except that for relatively light systems $a_4$ decreases to about 36 MeV.

While the typical value for the compressibility in the RMF-PC approach is $K
\approx 260-280$~MeV, this value is lowered to probably even more realistic
values in two adjustments, namely, the $\sigma$ fit and the fit only to masses with a percentage error.

\begin{table}[htb]
\begin{tabular}{l|ccccc}
Type of adjustment & $\rho_0$~[fm$^{-3}$] & E/A [MeV]& K [MeV] & $m^*/m$ & $a_4$ [MeV] \\\hline
PC-F1 & 0.151     & 16.17 & 270 & 0.61 & 37.8\\
PC-masses & 0.150 & 16.03 & 237 & 0.61 & 33.9 \\
PC-asym & 0.151 & 16.12 & 268 & 0.61 & 34.0 \\
PC-surf  & 0.147 & 16.28 & 238 & 0.62 & 40.6 \\
\end{tabular}
\caption{Bulk properties of nuclear matter for the various adjustment protocols discussed in the text.}
\label{tab1}
\end{table}

\section{Conclusions}
In this paper we have dealt with a less frequently discussed but nevertheless quite important topic in phenomenological nuclear structure, namely the adjustment
of the model parameters (coupling constants). We have studied the predictive power of the relativistic mean-field model with point-couplings
determined from various nuclear ground-state observables. The dependencies on the weights of different observables in the $\chi^2$ minimization and
on mass ranges of nuclei have been evaluated. Alternative adjustment protocols have been
tested and issues as well as features of current adjustment algorithms have been discussed.

We have tested various optimization algorithms and found the Levenberg-Marquardt method, supplemented with a trial step in parameter space (Bevington curved step),
to be the most efficient method. This is true for adjustments aiming at the nearest minimum. The $\chi^2$ hyper-surfaces of nuclear mean-field models,
however, contain several (local) minima. Thus, for adjustments of new forces, Monte-Carlo methods like simulated annealing are necessary
to explore the (unknown) parameter space. Once the various minima have been identified, a down-hill method can drive the parameter vector into the
minimum. 

The main outcome of these studies is that the present RMF forces are not able to describe simultaneously the mass and the geometry of the nucleus with the desired accuracy. 
The problem appears to be the surface thickness of the nucleus. It is underestimated in all forces under investigation, and this
underestimation is linked to the corresponding asymmetry energies. In the present sets, a larger and more
realistic surface thickness leads to an even larger value of $a_4$, which is already too high in RMF models ($a_4 \approx 38$~MeV)
when compared with values from liquid-drop mass fits yielding $a_4 \approx 32$~MeV.  This finding supports earlier results
on a correlation between $a_4$ and the neutron skin \cite{FurnstahlN}. 

Relativistic mean-field models are aimed at describing nuclei from $A\geq 16$ up to superheavy nuclei with $A \approx 300$.
Thus, the current mass range in our adjustments is from oxygen to lead isotopes. Our results show that the present forces constitute
a compromise of describing both light to medium and heavy systems. Heavy systems get over-bound if the adjustment
is confined to lighter nuclei. Correlations in lighter systems, which constitute a larger fraction of the binding energy
compared to heavy systems, demand stronger binding leading to too strong binding in heavy nuclei. Hence, current adjustments
have to live with this trade off. Possible solutions are more terms that can account for these missing correlations and/or
more phenomenological add-ons like a Wigner-energy term to account for missing $Z=N$ pairing.

More generally, our investigations indicate that the isoscalar and isovector channels are somewhat entangled in current forces and that hence the isovector channel might account for mismatches in the isoscalar part of the interaction. To avoid this entanglement, more observables 
of isovector type are needed, especially from nuclei with very large isospin.

It is clear that additional terms are needed to account for the missing physics in the present RMF approaches. 
They might -- to a large extent -- account for the missing correlations and density dependence.
Alternatively, more detailed range might be required.
A systematic exploration of various extensions in the framework of RMF-PC is underway \cite{EXT}.

\section*{Acknowledgments}
This work was supported by The U.S. Department of Energy
and by the Bundesministerium f\"ur Bildung und Forschung (BMBF),
Project No. 06 ER 124.

\end{document}